\documentclass[conference]{IEEEtran}
\IEEEoverridecommandlockouts

\usepackage{cite}
\usepackage{amsmath,amssymb,amsfonts}
\usepackage{algorithmic}
\usepackage{graphicx}
\usepackage{textcomp}
\usepackage{xcolor}
\def\BibTeX{{\rm B\kern-.05em{\sc i\kern-.025em b}\kern-.08em
 T\kern-.1667em\lower.7ex\hbox{E}\kern-.125emX}}
\begin{document}

\title{Bayesian network approach to building an affective module for a driver behavioural model
}

\author{\IEEEauthorblockN{1\textsuperscript{st} Dorota Młynarczyk\IEEEauthorrefmark{1}, 
2\textsuperscript{nd} Gabriel Calvo\IEEEauthorrefmark{1}, 
3\textsuperscript{rd} Francisco Palmi-Perales\IEEEauthorrefmark{1}, 
4\textsuperscript{th} Carmen Armero\IEEEauthorrefmark{1}\\
5\textsuperscript{th} Virgilio Gómez-Rubio\IEEEauthorrefmark{2}, 
6\textsuperscript{th} Ursula Martinez-Iranzo\IEEEauthorrefmark{3}}
\IEEEauthorblockA{\IEEEauthorrefmark{1}Departament d'Estadística i Investigació Operativa, Universitat de València, València, Spain}
\IEEEauthorblockA{\IEEEauthorrefmark{2}Departamento de Matemáticas, Universidad de Castilla-La Mancha, Albacete, Spain}
\IEEEauthorblockA{\IEEEauthorrefmark{3}Instituto de Biomecánica de València, Universitat Politècnica de València, València, Spain}}

\maketitle

\begin{abstract}
This paper focuses on the affective component of a Driver Behavioural Model (DBM), specifically modelling some driver's mental states, such as mental load and active fatigue, which may affect driving performance. We used Bayesian networks (BNs) to explore the dependencies between various relevant variables and estimate the probability that a driver was in a particular mental state based on their physiological and demographic conditions. Through this approach, our goal is to improve our understanding of driver behaviour in dynamic environments, with potential applications in traffic safety and autonomous vehicle technologies.\\
\end{abstract}

\section{Introduction}

Understanding driver behaviour is a cornerstone of improving road safety, optimising traffic management, and developing autonomous vehicle systems. Human drivers exhibit complex decision-making processes influenced by numerous factors, many of them with a high degree of uncertainty, including environmental conditions, traffic scenarios, and psychological states among others. To reflect human behaviour, a probabilistic model is required that can both account for its complexity and capture the uncertainty associated with it. Bayesian Networks (BNs) provide a robust framework to address this challenge.

A Bayesian Network (BN) \cite{Pearl1995} is a probabilistic model that uses a graphical framework to represent and analyse stochastic relationships between variables. In a BN, random variables are represented as nodes, and the edges between them encode probabilistic conditional dependencies. The strength of these relationships is quantified via conditional probability distributions, allowing the model to assess probabilities associated to different outcomes based on observed data. From a modelling perspective, BNs offer several advantages, such as their ability to represent uncertain relationships, model complex dependencies, provide a flexible framework for inference, and as \cite{pearl1995bayesian} state ``they are direct representations of the world, not of reasoning processes''. By incorporating probabilistic reasoning, BNs are a powerful tool for modelling complex systems. This approach is particularly suitable for analysing driver behaviour, as it can effectively accommodate the unpredictability inherent in human decision-making processes \cite{forbes1995batmobile}.

A Driver Behavioural Model (DBM) \cite{Negash2023} is a framework that integrates data from various sources, such as sensors, traffic systems, or driver profiles, to study how different factors (such as speed, road conditions, and distractions) influence driver actions. DBMs can be used to improve road safety, support driver assistance systems, or design autonomous vehicles.

In this paper, we focus on the affective module of a DBM, which is designed to capture and model the emotional and cognitive states that influence the behaviour of a driver. Specifically, our objective is to examine the intensity of some mental states, such as active fatigue and mental load, that significantly impact on how a driver responds to various traffic scenarios. The affective module integrates data from physiological indicators, such as heart rate and respiration rate, to assess these mental states. Furthermore, the model presented in this paper can be expanded to explore how mental states influence driving actions, offering a more comprehensive understanding of driver behaviour under dynamic and uncertain conditions.

The remainder of this paper is organised as follows. Section~\ref{sec:BN} outlines the foundational principles of Bayesian Belief Networks. Section~\ref{sec:BNAM} provides a brief review of Bayesian Belief Networks in affective models. Section~\ref{sec:BBN} details the database for the model and presents an example of the network. Finally, Section~\ref{sec:discussion} briefly discusses the extension of the model and areas for further research.\\

\section{Bayesian networks}
\label{sec:BN}
 According to \cite{pearl1995bayesian}, a BN is a joint probability distribution over a set of random variables $\boldsymbol{Y} = \{Y_v : v \in \mathcal{V} \}$, where $\mathcal{V} = \{1,2,\dots,V\}$ represents the node set of the network. That joint distribution is denoted as
\begin{align*}
 f(\boldsymbol{y}).
\end{align*}

\noindent Each node $Y_v$ has a set of parent nodes, denoted as $\operatorname{pa}(Y_v)$, which consists of variables that have directed edges pointing to $Y_v$ in the network structure. The dependency of $Y_v$ on its parents is expressed through the conditional distribution $f(y_v \mid y_{\operatorname{pa}(Y_v)})$. If $Y_v$ has no parents, its distribution is given by its marginal probability distribution $f(y_v)$. The children of node $Y_v$, $\operatorname{ch}(Y_v)$, are all nodes that have direct arrows to it. The family of a node is itself and its parents.

The factorization property (also known as global semantics) of a Bayesian network states that the joint probability distribution decomposes in terms of univariate conditional distributions as follows
\begin{align*}
 f(\boldsymbol{y}) = \prod_{v=1}^{V} f(y_v \mid y_{\operatorname{pa}(Y_v)}).
\end{align*}
\noindent This factorization encodes the conditional dependence assumptions implied by the network structure. 

BN models can always be represented as graphs, where each node corresponds to a random variable and each edge of the graph represents a probability distribution that connects a parent variable to one of its children. These graphs must adhere to two general properties: all edges must have a direction, and the graph must be acyclic. These are called directed acyclic graphs (DAG).

\begin{figure}[htbp]
\centerline{\includegraphics[width=4cm]{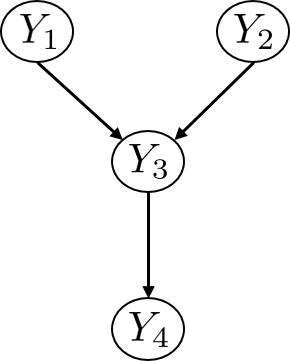}}
\caption{Basic direct acyclic graph example.}\label{figure_1}
\end{figure}

We illustrate these concepts by means of a toy BN with nodes $\mathcal{V}=\{1,2,3,4\}$ and random variables $\boldsymbol{Y}=\{Y_1, Y_2, Y_3, Y_4\}$ with joint probability distribution $$ f(\boldsymbol y) = f(y_4| y_3)f(y_3| y_1, y_2)f(y_1)f(y_2).$$ \noindent This model is represented as a directed acyclic graph (DAG), as shown in Figure \ref{figure_1}. On the one hand, we can observe that both $Y_1$ and $Y_2$ are parents of $Y_3$. Moreover, neither $Y_1$ nor $Y_2$ has parents, so their distributions are given by the marginal distributions $f(y_1)$ and $f(y_2)$, respectively. Finally, we can see that the variable $Y_4$ is independent of $Y_1$ and $Y_2$, given its parent $Y_3$.

Dynamic Bayesian Networks (DBNs) extend the framework of static BNs by introducing a temporal dimension into the network structure \cite{dean1989model}. This enhancement allows DBNs to model time-varying systems more effectively, capturing the evolution of variables and their dependencies over time. While DBNs are particularly well-suited to systems with temporal or sequential dynamics, the inclusion of temporal or spatial elements significantly increases the complexity of both their structure and analysis.

Finally, it is worth highlighting that, in general, two statistical strategies can be followed to estimate the values of the parameters $\boldsymbol{\theta}$ involved in the probabilistic model \cite{Wood2015}. One option is the Bayesian approach, which models uncertainty in probabilistic terms and requires assigning a prior distribution to $\boldsymbol{\theta}$ which models uncertainty in probabilistic terms and requires a prior probability distribution to initiate the inferential process, that will be subsequently updated with the input of information provided by the data. The alternative statistical approach, which we have adopted in this work, is to estimate the parameter values of $\boldsymbol{\theta}$ from a frequentist perspective. Specifically, we have estimated them by computing the maximum likelihood values based on the available observations. We include all the variables in the network because their correlations are not very high and, consequently, we do not seem to have multicollinearity problems between them.\\

\section{BN in affective models}
\label{sec:BNAM}
Despite the growing interest in introducing the study of stochastic variability in autonomous vehicle driving problems, the authors of \cite{wang2023quantification} highlight the lack of quantifiable data and uncertainty of the modelling in this field, thus emphasizing the enormous potential of BNs for application in such problems. In particular, there are interesting studies that apply this type of models to analyse how affective or mental states influence driver behaviour in various situations. An example can be found in \cite{liu2020analysis}, who developed a BN to analyse how emotions affect a driver's behaviour. Specifically, it identifies eight emotional states: anger, anxiety, contempt, fear, helplessness, pleasure, relief, and surprise.

However, in our study, our primary objective is to capture the mental state or states of a driver, based primarily on various physiological indicators, such as blood pressure and respiration. Some studies follow this direction, such as \cite{he2015driver}, where the authors use BNs to analyse driver fatigue based on indicators such as electroencephalogram (EEG) signals and head nodding angles. Notably, they utilise latent variables effectively. In their approach, the variables of interest—namely, alertness and drowsiness—are appropriately hidden and informed by the observed indicators and contextual variables. A similar approach is presented in \cite{yang2010driver}, where the authors developed a DBN to capture driver fatigue by integrating contextual factors (e.g., sleep quality, work environment) and physiological signals (e.g., EEG, ECG). By employing a first-order Hidden Markov Model, the DBN tracks changes over time, enabling a dynamic assessment of fatigue.\\

\section{Bayesian Belief Networks for affective module of the DBM model}
\label{sec:BBN}
\subsection{Dataset}
Data for this paper comes from a study conducted by the Instituto de Biomecánica de Valencia. The dataset includes a broad range of information, encompassing demographic, biological, and physiological data about the participants, along with responses to experimental questionnaires designed to assess various mental states, such as mental load and active fatigue. The study was designed to include both men and women across a broad age range, as other studies suggest that there are certain differences in driving behaviours based on sex and age \cite{rhodes2011age}. 

The dataset comprises 1,892 observations collected from 56 participants, with the following basic information:
\begin{itemize}
 \item \textbf{Sex}: 28 males and 28 females. \vspace*{0.06cm}
 \item \textbf{Age}: Ranging from 21 to 61 years (median: 45.5 years; mean: 44.06 years).\vspace*{0.05cm}
 \item \textbf{Weight}: Ranging from 49 kg to 115 kg (median: 71.5 kg; mean: 71.66 kg). \vspace*{0.05cm}
 \item \textbf{Height}: Ranging from 156 cm to 192 cm (median: 171 cm; mean: 170.9 cm). \vspace*{0.05cm}
 \item \textbf{Body Mass Index (BMI)}: Ranging from 17.65 to 36.71 (median: 23.20; mean: 24.46). \vspace*{0.05cm}
\end{itemize}

\begin{figure}[htbp]
\centerline{\includegraphics[width=6cm]{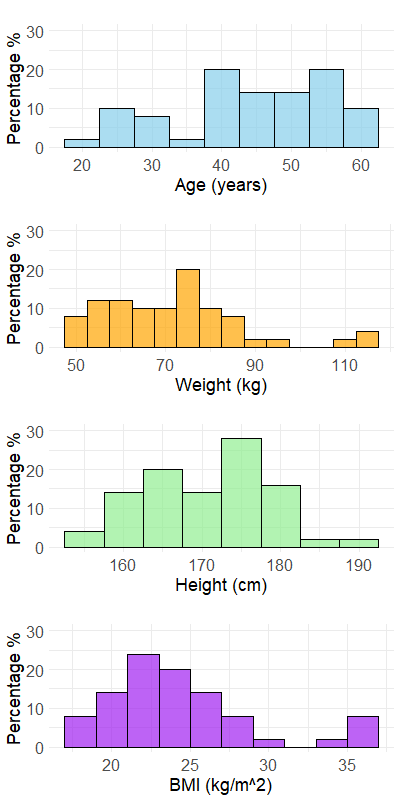}}
\caption{Histograms illustrating the basic characteristics of the study participants.}\label{figure_2}
\end{figure}

Histograms in Figure \ref{figure_2} show the distribution of the participant data. They evidence the variability of the selected sample.
\indent Physiological assessment was performed allowing for the inclusion of the following variables: 

\begin{itemize} 
 \item \textbf{SDNN:} Standard deviation of the time elapsed between two successive normal heartbeats (RR intervals).\vspace*{0.05cm} 
 
 \item \textbf{SDSD:} Standard deviation of the successive differences between adjacent RR intervals. \vspace*{0.05cm} 
 
 \item \textbf{Mean\_HR:} (Mean Heart Rate) 
 Average heart rate, indicating the mean number of beats per minute over a specific period. \vspace*{0.05cm} 
 
 \item \textbf{LF\_HF\_ratio:} Ratio of the variance of the Heart Rate Variability (HRV) in the low-frequency band (0.04–0.15 Hz) to the high-frequency band (0.15–0.40 Hz). \vspace*{0.05cm} 
 
 \item \textbf{Resp\_rate:} 
 Mean of the number of breaths per minute. \vspace*{0.05cm}
\end{itemize}
Therefore, the dataset includes four variables related to heart function and one related to respiration. The correlations among these physiological variables can be seen in Table \ref{tab:correlation_matrix}. We include all the variables in the network because their correlations are not very high and, consequently, we do not seem to have multicollinearity problems between them.

\begin{table}[htbp]
\caption{Correlation matrix of the physiological variables}
\begin{center}
\resizebox{0.48\textwidth}{!}{
\begin{tabular}{|c|c|c|c|c|c|}
\hline
\textbf{} & \textbf{SDNN} & \textbf{SDSD} & \textbf{Mean\_HR} & \textbf{LF\_HF\_ratio} & \textbf{Resp\_rate} \\
\hline
\textbf{SDNN} & 1.000 & 0.401 & 0.412 & 0.320 & -0.305 \\
\textbf{SDSD} & 0.401 & 1.000 & -0.217 & -0.292 & -0.064 \\
\textbf{Mean\_HR} & 0.412 & -0.217 & 1.000 & 0.175 & -0.177 \\
\textbf{LF\_HF\_ratio} & 0.320 & -0.292 & 0.175 & 1.000 & -0.136 \\
\textbf{Resp\_rate} & -0.305 & -0.064 & -0.177 & -0.136 & 1.000 \\
\hline
\end{tabular}}
\label{tab:correlation_matrix}
\end{center}
\end{table}

Following the experimental tasks, participants completed a series of mental state questionnaires. The original responses were recorded on a continuous scale ranging from 0 to 100. However, for this analysis, the responses were transformed into a binary format with two distinct outcomes: 1, indicating that the participant entered the specified mental state, and 0, indicating otherwise. The mental states considered in this study were mental load (\textbf{ML}) and active fatigue (\textbf{AF}).

\subsection{BBN}
This Bayesian network was fitted using the R package \texttt{bnlearn} \cite{scutari2010bnlearn}, a popular tool that includes several algorithms for learning the structure of Bayesian networks. The network was built by exploring multiple possible configurations using algorithms that estimate dependencies between variables based on the data. Once the models were generated, the best-fitting network was selected based on the Bayesian Information Criterion (BIC) \cite{neath2012bic}. BIC provides a robust method for model comparison by balancing goodness of fit and model complexity.

\begin{figure}[htbp]
\centerline{\includegraphics[width=9cm]{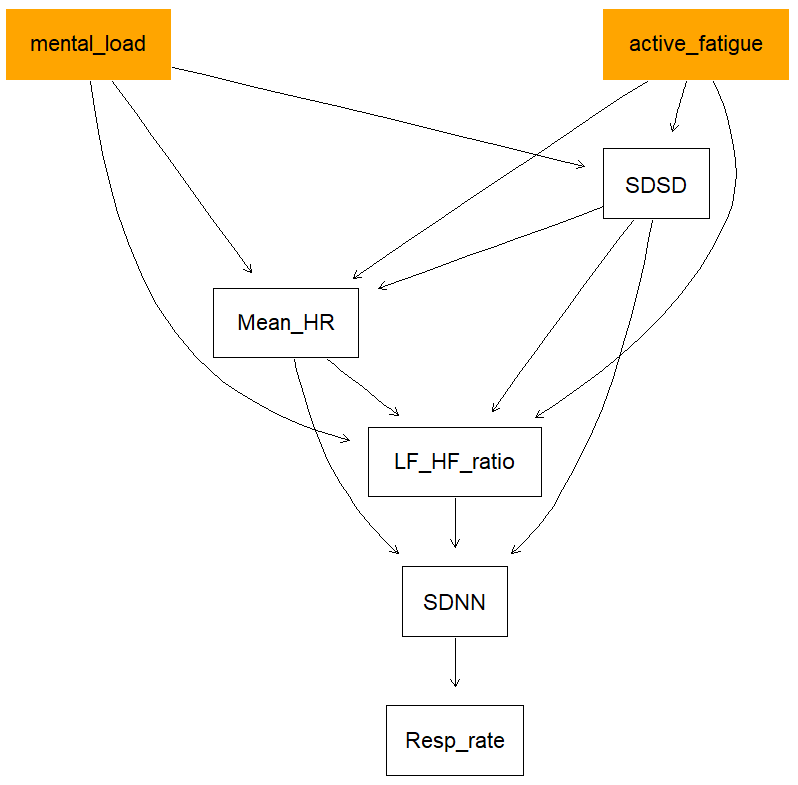}}
\caption{Bayesian Network for two mental states (coloured in orange).}\label{figure_3}
\end{figure}

Figure \ref{figure_3} presents the DAG of the selected Bayesian network. It illustrates the relationships among mental load (ML), active fatigue (AF), and the set of physiological variables: SDNN, SDSD, Mean\_HR, LF\_HF\_ratio, and Resp\_rate. For transparency, the covariates are not included in the presented model. The subsequent joint distribution can be decomposed as the product
\[
\begin{aligned}
f(\boldsymbol{y}) &= f(\text{Resp\_rate} \mid \text{SDNN}) \cdot \\
&\quad f(\text{SDNN} \mid \text{SDSD}, \text{Mean\_HR}, \text{LF\_HF\_ratio}) \cdot \\
&\quad f(\text{LF\_HF\_ratio} \mid \text{SDSD}, \text{Mean\_HR}, \text{ML}, \text{AF}) \cdot \\
&\quad f(\text{Mean\_HR} \mid \text{SDSD}, \text{ML}, \text{AF})  \cdot\\
&\quad f(\text{SDSD} \mid \text{ML}, \text{AF}) f(\text{ML}) f(\text{AF}).
\end{aligned}
\]

In this network, mental load and active\_fatigue have a direct influence on Mean\_HR, SDSD, and LF\_HF\_ratio. SDSD affects Mean\_HR, and both SDSD and Mean\_HR contribute to the LF\_HF\_ratio. These three variables (Mean\_HR, SDSD, LF\_HF\_ratio) influence SDNN, which in turn affects Resp\_rate. These interdependencies illustrate the complex relationships within the network. In terms of interpreting these variables, it can be stated that, according to the model, mental states influence heart function, which in turn affects respiration.

After defining the geometry of the BN, we can model the conditional distribution of all nodes. For instance, the conditional distribution of the variable Mean\_HR, given that SDSD, ML, and AF are its parent variables, follows a normal distribution with parameters $\mu$ (mean) and $\sigma$ (standard deviation): 

\[
f(\text{Mean\_HR} \mid \text{SDSD}, \text{ML}, \text{AF}) = \mathcal{N}(\mu, \sigma^2).
\] 

\begin{figure}[htbp]
\centerline{\includegraphics[width=9cm]{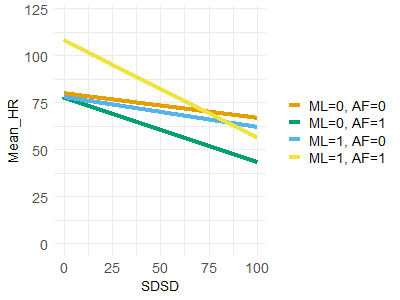}}
\caption{Mean ($\mu$) of the conditional distribution of the Mean\_HR variable with regard to SDSD for the four possible scenarios defined by  ML and AF.}\label{figure_4}
\end{figure}

The parameters $\mu$ and $\sigma$ depend on the different possible values of the two binary mental state variables, ML and AF. These cases include scenarios where both variables are zero, when one of the variables is one (two separate cases), and when both variables are one. Results are presented in Table \ref{tab:mean_hr}, where these four cases are represented as rows. Figure \ref{figure_4} illustrates the mean $\mu$ of the conditional distribution, with separate lines representing the different scenarios.

\begin{table}[htbp]
\caption{Estimates of the parameters of the conditional distribution of the Mean\_HR variable given SDSD, ML and AF}
\label{tab:mean_hr}
\centering
\resizebox{0.48\textwidth}{!}{
\begin{tabular}{|c|c|c|}
\hline
\textbf{Case} & \textbf{$\mu$} & \textbf{$\sigma$} \\
\hline
\textbf{ML=0, AF=0} & $79.930 - 0.130 \cdot \text{SDSD}$ & 13.654 \\
\textbf{ML=1, AF=0} & $77.967 - 0.159 \cdot \text{SDSD}$ & 14.879 \\
\textbf{ML=0, AF=1} & $77.670 - 0.342 \cdot \text{SDSD}$ & 4.432 \\
\textbf{ML=1, AF=1} & $108.161 - 0.516 \cdot \text{SDSD}$ & 26.755 \\
\hline
\end{tabular}
}
\end{table}

The regression coefficients of SDSD in  Table \ref{tab:mean_hr}  show a negative correlation between this variable and Mean\_HR, meaning that as SDSD increases, Mean\_HR tends to decrease. The difference in slopes indicates that SDSD affects active fatigue (AF=1) and non-active fatigue (AF=0) instances differently; cases with AF=1 (green and yellow lines) have steeper slopes compared to non-AF situations, suggesting a stronger effect of SDSD on Mean\_HR in active fatigue conditions.

The estimated model also allows us to assess the probability that somebody is in a particular mental state given the values of a subset of the other variables. For example, the probability that a person is under mental load given that their mean heart rate is over 100 and their respiration rate is greater than 20 (per minute) is equal to 94\%
\[Prob\,(\text{ML}=1 \mid \text{Mean\_HR}>100, \text{Resp\_rate}>20)=0.94.\]
For the same conditions, the probability that a person has active fatigue is estimated to be 78\%\[
Prob\,(\text{AF}=1 \mid \text{Mean\_HR}>100, \text{Resp\_rate}>20)=0.78\]

The probability that a person is in one of four possible situations, each involving a combination of two mental states, given that their mean heart rate exceeds 100 beats per minute and their respiration rate is greater than 20 breaths per minute, is presented in Table \ref{tab:prob}. Under these conditions, the most probable scenario is that the person is experiencing mental load and active fatigue. The events where mental load is absent have very low probabilities; in fact, the probability associated to the case ML = 0 and AF = 1 is exactly 0.

\begin{table}[htbp]
\caption{Probability that a person is in mental states given that mean heart rate is over 100 and respiration rate is greater than 20}
\begin{center}
\resizebox{0.25\textwidth}{!}{
\begin{tabular}{|c|c|}
\hline
\textbf{Case} & \textbf{Probability} \\
\hline
\textbf{ML=0, AF=0} & 0.057\\
\textbf{ML=1, AF=0} & 0.161 \\
\textbf{ML=0, AF=1} & 0.000 \\
\textbf{ML=1, AF=1} & 0.782 \\
\hline
\end{tabular}}
\label{tab:prob}
\end{center}
\end{table}

\section{Discussion}
\label{sec:discussion}
The ability to estimate the probability of being in a specific mental state, such as the likelihood of mental load or active fatigue under certain physiological conditions, is one of the most important aspects of this study. This could allow for real-time assessments of the driver’s mental state, which has some implications for both traffic safety and autonomous vehicle technology. For instance, a vehicle could alert the driver when some signs of mental fatigue are detected, suggesting that they  should take a break. The application of this model could enhance the design of systems aimed at preventing accidents caused by cognitive fatigue or overload.

While the results of this study show promising applications, there are also areas for further research. For example, while we focused on a limited set of physiological variables in the current model, it may be beneficial to explore additional markers, such as facial expressions, which could provide further insights into mental states. Moreover, the model could be extended to incorporate environmental factors, such as road conditions and traffic density, that may also impact driver perception.

\section*{Acknowledgment}
Research activity under BERTHA project (GA101076360) funded by the European Union. Views and opinions expressed are however those of the author(s) only and do not necessarily reflect those of the European Union or the European Climate, Infrastructure and Environment Executive Agency (CINEA). Neither the European Union nor the granting authority can be held responsible for them.\\

\end{document}